\documentclass[aip,jcp,preprint,amsmath,amsfonts,amssymb,showkeys]{revtex4-1}
\usepackage{graphicx,dcolumn,bm,xcolor,microtype,hyperref,physics,float}
\usepackage[version=4]{mhchem}

\newcommand{\Fx}{F_{\text{x}}}
\newcommand{\gX}{gX}

\newcommand{\GX}{GX}

\newcommand{\PBEGX}{PBE-GX}
\newcommand{\FxgX}{F_{\text{x}}^\text{\gX}}

\newcommand{\FxGX}{F_{\text{x}}^\text{\GX}}

\newcommand{\FxPBEGX}{F_{\text{x}}^\text{\PBEGX}}

\newcommand{\FxGGA}{F_{\text{x}}^\text{GGA}}
\newcommand{\FxGLDA}{F_{\text{x}}^\text{GLDA}}
\newcommand{\FxMGGA}{F_{\text{x}}^\text{MGGA}}
\newcommand{\FxFMGGA}{F_{\text{x}}^\text{FMGGA}}

\newcommand{\exsLDA}{e_{\text{x},\sigma}^\text{LDA}}
\newcommand{\exsGGA}{e_{\text{x},\sigma}^\text{GGA}}
\newcommand{\exsGLDA}{e_{\text{x},\sigma}^\text{GLDA}}
\newcommand{\exsMGGA}{e_{\text{x},\sigma}^\text{MGGA}}
\newcommand{\exsFMGGA}{e_{\text{x},\sigma}^\text{FMGGA}}
\newcommand{\CxLDA}{C_\text{x}^\text{LDA}}
\newcommand{\CxGLDA}{C_\text{x}^\text{GLDA}}
\newcommand{\Cx}{C_\text{x}}
\newcommand{\Cf}{C_\text{F}}
\newcommand{\rs}{\rho_\sigma}
\newcommand{\xs}{x_\sigma}

\newcommand{\as}{\alpha_\sigma}
\newcommand{\ns}{n_\sigma}
\newcommand{\Ls}{L_\sigma}
\newcommand\upa{\uparrow}
\newcommand\dwa{\downarrow}
\newcommand{\br}{\bm{r}}
\newcommand{\mc}{\multicolumn}

\newcommand{\alert}[1]{\textcolor{black}{#1}}

\usepackage{tikz}
\usetikzlibrary{arrows,positioning,shapes.geometric}
\usetikzlibrary{decorations.pathmorphing}

\begin{document}

\title{Exchange functionals based on finite uniform electron gases}

\author{Pierre-Fran{\c c}ois Loos}
\email{loos@irsmac.ups-tlse.fr}
\affiliation{Laboratoire de Chimie et Physique Quantiques, Universit\'e de Toulouse, CNRS, UPS, France}
\affiliation{Research School of Chemistry, Australian National University, Canberra ACT 2601, Australia}

\begin{abstract}
We show how one can construct \alert{a simple} exchange functional by extending the well-know local-density approximation (LDA) to finite uniform electron gases.
This new generalized local-density approximation (GLDA) functional uses only two quantities: the electron density $\rho$ and the curvature of the Fermi hole $\alpha$.
This alternative ``rung 2'' functional can be easily coupled with generalized-gradient approximation (GGA) functionals to form a new family of ``rung 3'' meta-GGA (MGGA) functionals that we have named factorizable MGGAs (FMGGAs).
Comparisons are made with various LDA, GGA and MGGA functionals for atoms and molecules.
\end{abstract}

\keywords{exchange functional; local-density approximation; generalized local-density approximation; density-functional theory}

\maketitle

\section{
\label{sec:intro}
Introduction}
\begin{figure}
	\includegraphics[width=0.49\linewidth]{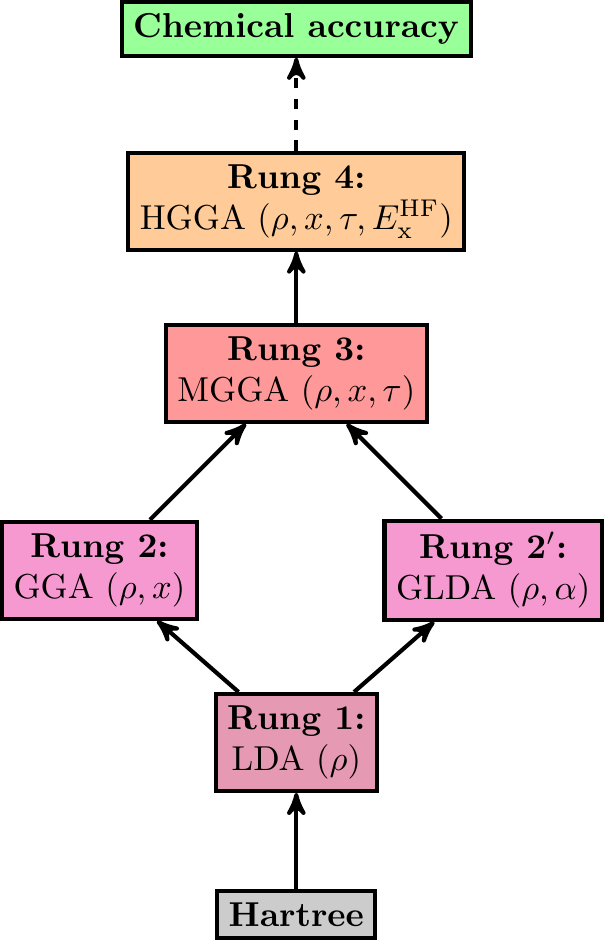}
\caption{
\label{fig:Jacob}
Jacob's ladder of DFT revisited. $\rho$, $x$, $\tau$, $\alpha$ and $E_\text{x}^\text{HF}$ are the electron density, the reduced gradient, the kinetic energy density, the curvature of the Fermi hole and the Hartree-Fock (HF) exchange energy, respectively.
The fourth rung corresponds to hyper GGA (HGGA) functionals.}
\end{figure}

Due to its moderate computational cost and its reasonable accuracy, Kohn-Sham (KS) density-functional theory \cite{Hohenberg64, Kohn65} (DFT) has become the workhorse of electronic structure calculations for atoms, molecules and solids. \cite{ParrBook}
To obtain accurate results within DFT, one only requires the exchange and correlation functionals, which can be classified in various families depending on their physical input quantities. \cite{Becke14, Yu16}
These various types of functionals are classified by the Jacob's ladder of DFT \cite{Perdew01, Pewdew05} (see Fig.~\ref{fig:Jacob}).
The local-density approximation (LDA) sits on the first rung of the Jacob's ladder and only uses as input the electron density $\rho$. 
The generalized-gradient approximation (GGA) corresponds to the second rung and adds the gradient of the electron density $\nabla \rho$ as an extra ingredient. 
The third rung is composed by the so-called meta-GGA (MGGA) functionals \cite{Sala16} which uses, in addition to $\rho$ and $\nabla \rho$, the kinetic energy density $\tau = \sum_i^\text{occ} \abs{\nabla\psi_{i} }^2$ (where $\psi_i$ is an occupied molecular orbital).

The infinite uniform electron gas (IUEG) or jellium \cite{1DEG13, 2DEG11, 3DEG11, Handler12, WIREs16} is a much studied and well-understood model system, and hence a logical starting point for local exchange-correlation approximations. \cite{VWN, PZ81, Perdew81, Becke83,  Becke86, Becke88b, Wang91, PW92, Sun10}
Though analytical models are scarce, we have recently discovered an entire new family of analytical models that one can use to develop new exchange and correlation functionals within DFT. \cite{UEGs12, Ringium13, gLDA14, Wirium14, LowGlo15, SBLDA16, ESWC17} 
Indeed, we have shown that, by constraining $n$ electrons on a surface of a three-dimensional sphere (or 3-sphere), one can create finite uniform electron gases (FUEGs). \cite{Glomium11, LowGlo15, WIREs16}
Here, we show how to use these FUEGs to create a new type of exchange functionals applicable to any type of systems.
We have already successfully applied this strategy to one-dimensional systems, \cite{1DChem15, SBLDA16, Leglag17} for which we have created a correlation functional based on this idea. \cite{gLDA14, Wirium14}
Moreover, we show that these alternative second-rung functionals can be easily coupled to GGA functionals to form a new family of third-rung MGGA functionals. 
Unless otherwise stated, we use atomic units throughout.

\section{
\label{sec:theory}
Theory}
Within DFT, one can write the total exchange energy
as the sum of its spin-up ($\sigma=$ $\upa$) and spin-down ($\sigma=$ $\dwa$) contributions:
\begin{equation}
	E_\text{x} = E_{\text{x},\upa} + E_{\text{x},\dwa},
\end{equation}
where
\begin{equation}
	E_{\text{x},\sigma} = \int e_{\text{x},\sigma}(\rho_\sigma,\nabla \rho_\sigma, \tau_\sigma, \ldots) \, \rho_\sigma(\br) \, d\br,
\end{equation}
and $\rs$ is the electron density of the spin-$\sigma$ electrons.
\alert{Although, for sake of simplicity, we sometimes remove the subscript $\sigma$, we only use spin-polarized quantities from hereon.}

The first-rung LDA exchange functional (or D30 \cite{Dirac30}) is based on the IUEG \cite{WIREs16} and reads
\begin{equation}
	\exsLDA(\rs) = \CxLDA \rs^{1/3},
\end{equation}
where
\begin{equation}
	\CxLDA = - \frac{3}{2} \qty( \frac{3}{4\pi} )^{1/3}.
\end{equation}

A GGA functional (second rung) is defined as
\begin{equation}
	\label{eq:GGA}
	\exsGGA(\rs,\xs) 
	= e_{\text{x},\sigma}^\text{LDA}(\rs)  \FxGGA(\xs),
\end{equation}
where $\FxGGA$ is the GGA enhancement factor depending only on the reduced gradient
\begin{equation}
	x = \frac{\abs{\nabla \rho}}{\rho^{4/3}},
\end{equation}
and
\begin{equation}
	\lim_{x \to 0} \FxGGA(x) = 1,
\end{equation}
i.e.~a well thought-out GGA functional reduces to the LDA for homogeneous systems.
The well-known B88, G96, PW91 and PBE exchange functionals are examples of GGA functionals. \cite{B88, G96, PW91, PBE}

\alert{Similarly, motivated by the work of Becke \cite{Becke00} and our previous investigations, \cite{gLDA14, Wirium14}} we define an alternative second-rung functional that we call generalized LDA (GLDA) 
\begin{equation}
\label{eq:GLDA}
	\exsGLDA(\rs,\as) = \exsLDA(\rs)  \FxGLDA(\as).
\end{equation}

By definition, a GLDA functional only depends on the electron density and the curvature of the Fermi hole (see Fig.~\ref{fig:Jacob}):
\begin{equation} 
\label{eq:eta-def}
        \alpha = \frac{\tau - \tau_\text{W}}{\tau_\text{IUEG}} = \frac{\tau}{\tau_\text{IUEG}} - \frac{x^2}{4 \Cf},
\end{equation}
which measures the tightness of the exchange hole around an electron. \cite{Becke83, Dobson91}
In Eq.~\eqref{eq:eta-def},
\begin{equation}
	\tau_\text{W} = \frac{\abs{\nabla\rho}^2}{4\,\rho}
\end{equation}
is the von Weizs{\"a}cker kinetic energy density, \cite{vonWeizsacker35} and 
\begin{equation}
	\tau_\text{IUEG} = \Cf \rho^{5/3} 
\end{equation}
is the kinetic energy density of the IUEG, \cite{WIREs16} where
\begin{equation}
	\Cf = \frac{3}{5} (6\pi^2)^{2/3}.
\end{equation}
The dimensionless parameter $\alpha$ has two characteristic features: i) $\alpha=0$ for any one-electron system, and ii) $\alpha=1$ for the IUEG.
Some authors call $\alpha$ the inhomogeneity parameter but we will avoid using this term as we are going to show that $\alpha$ can have distinct values in homogeneous systems.
For well-designed GLDA functionals, we must ensure that
\begin{equation}
\label{eq:limGLDA}
	\lim_{\alpha \to 1} \FxGLDA(\alpha) = 1,
\end{equation}
i.e.~the GLDA reduces to the LDA for the IUEG.\footnote{While some functionals only use the variable $\tau$, \cite{Ernzerhof99, Eich14} we are not aware of any functional only requiring $\alpha$.}

Although any functional depending on the reduced gradient $x$ and the kinetic energy density $\tau$ is said to be of MGGA type, here we will define a third-rung MGGA functional as depending on $\rho$, $x$ and $\alpha$:
\begin{equation}
	\exsMGGA(\rs,\xs,\as) = \exsLDA(\rs) \FxMGGA(\xs,\as),
\end{equation}
where one should ensure that
\begin{equation}
\label{eq:limMGGA}
	\lim_{x \to 0} \lim_{\alpha \to 1} \FxMGGA(x,\alpha) = 1,
\end{equation}
i.e.~the MGGA reduces to the LDA for an infinite homogeneous system.
\alert{The M06-L functional from Zhao and Truhlar, \cite{M06L} the mBEEF functional from Wellendorff et al. \cite{mBEEF} and the SCAN \cite{SCAN} and MS \cite{MS0,MS1_MS2} family of functionals from Sun et al.~are examples of widely-used MGGA functionals.}

The Fermi hole curvature $\alpha$ has been shown to be a better variable than the kinetic energy density $\tau$ as one can discriminate between covalent ($\alpha=0$), metallic ($\alpha \approx 1$) and weak bonds ($\alpha \gg 0$). \cite{Kurth99, TPSS, revTPSS, MS0, Sun13a, MS1_MS2, MVS, SCAN, Sun16b} 
The variable $\alpha$ is also related to the electron localization function (ELF) designed to identify chemical bonds in molecules. \cite{Becke83, ELF}
Moreover, by using the variables $x$ and $\alpha$, we satisfy the correct uniform coordinate density-scaling behavior. \cite{Levy85}

In conventional MGGAs, the dependence in $x$ and $\alpha$ can be strongly entangled, while, in GGAs for example, $\rho$ and $x$ are strictly disentangled as illustrated in Eq.~\eqref{eq:GGA}.
Therefore, it feels natural to follow the same strategy for MGGAs.
Thus, we consider a special class of MGGA functionals that we call factorizable MGGAs (FMGGAs)
\begin{equation}
	\exsFMGGA(\rs,\xs,\as) = \exsLDA(\rs) \FxFMGGA(\xs,\as),
\end{equation}
where the enhancement factor is written as 
\begin{equation}
	\FxFMGGA(x,\alpha) = \FxGGA(x) \FxGLDA(\alpha).
\end{equation}
By construction, $\FxFMGGA$ fulfills Eq.~\eqref{eq:limMGGA} and the additional physical limits
\begin{subequations}
\begin{align}
	\lim_{x \to 0} \FxFMGGA(x,\alpha) & = \FxGLDA(\alpha),
	\\
	\lim_{\alpha \to 1} \FxFMGGA(x,\alpha) & = \FxGGA(x).
\end{align}
\end{subequations}
The MVS functional designed by Sun, Perdew and Ruzsinszky is an example of FMGGA functional. \cite{MVS}

\section{
\label{sec:functional}
Exchange functionals}

\subsection{
\label{sec:compdetails}
Computational details}
Unless otherwise stated, all calculations have been performed self-consistently with a development version of the Q-Chem4.4 package \cite{qchem4} using the aug-cc-pVTZ basis set. \cite{Dunning89, Kendall92, Woon93, Woon94, Woon95, Peterson02} 
To remove quadrature errors, we have used a very large quadrature grids consisting of 100 radial points (Euler-MacLaurin quadrature) and 590 angular points (Lebedev quadrature).
As a benchmark, we have calculated the (exact) unrestricted Hartree-Fock (UHF) exchange energies.

\subsection{
\label{sec:GLDA}
GLDA exchange functionals}

The orbitals for an electron on a 3-sphere of unit radius are the normalized hyperspherical harmonics $Y_{\ell\mu}$, where $\ell$ is the principal quantum number and $\mu$ is a composite index of the remaining two quantum numbers. \cite{AveryBook, Avery93}  
We confine our attention to ferromagnetic (i.e.~spin-polarized) systems in which each orbital with $\ell = 0, 1, \ldots , L_{\sigma}$ is occupied by one spin-up or spin-down electron, thus yielding an electron density that is uniform over the surface of the sphere.
Note that the present paradigm is equivalent to the jellium model \cite{WIREs16} for $L_{\sigma} \to \infty$.
We refer the reader to Ref.~\onlinecite{Glomium11} for more details about this paradigm.

The number of spin-$\sigma$ electrons is
\begin{equation}
	\ns = \frac{1}{3} (\Ls+1)(\Ls+3/2)(\Ls+2),
\end{equation}
\alert{and their one-electron uniform density around the 3-sphere is
\begin{equation}
	\rs = \frac{\ns}{V} = \frac{(\Ls+2)(\Ls+3/2)(\Ls+1)}{6 \pi^2 R^3},
\end{equation}
where $V = 2 \pi^2 R^3$ is the surface of a 3-sphere of radius $R$.
Moreover, using Eq.~\eqref{eq:eta-def}, one can easily derive that \cite{gLDA14, Wirium14}
\begin{equation}
\label{eq:alpha}
	\as = \frac{\Ls(\Ls+3)}{\qty[ (\Ls+1)(\Ls+3/2)(\Ls+2) ]^{2/3}},
\end{equation}}
which yields
\begin{align}
	\lim_{\ns \to 1 } \as & = 0,
	& 
	\lim_{\ns \to \infty } \as & = 1.
\end{align}
We recover the results that $\alpha = 0$ in a one-electron system (here a one-electron FUEG), and that $\alpha = 1$ in the IUEG.

\alert{In particular, we have shown that the exchange energy of these systems can be written as \cite{Glomium11, Jellook12}
\begin{equation}
\label{eq:Ex}
	E_{\text{x},\sigma}(\Ls) = \Cx(\Ls) \int \rs^{4/3} d\bm{r}.
\end{equation}
where
\begin{equation}
\label{eq:CxGLDA}
	\Cx(L) 
	= \CxLDA \frac{\frac{1}{2} \qty( L+\frac{5}{4}) \qty(L+\frac{7}{4}) \qty[\frac{1}{2} H_{2 L+\frac{5}{2}} + \ln 2] 
	+ \qty(L+\frac{3}{2})^2 \qty(L^2+3 L+\frac{13}{8})}
	{\qty[(L+1) \qty(L+\frac{3}{2}) (L+2)]^{4/3}}
\end{equation}
and $H_{k}$ is an harmonic number. \cite{NISTbook}}

\alert{Therefore, thanks to the one-to-one mapping between  $\Ls$ and $\as$ evidenced by Eq.~\eqref{eq:alpha}, we have created the \gX~functional}
\begin{equation}
\label{eq:FxGLDA}
	\FxgX(\alpha) = \frac{\CxGLDA(0)}{\CxGLDA(1)}
	\\
	+ \alpha \frac{c_0+c_1\,\alpha}{1+(c_0+c_1-1)\alpha} \qty[ 1 - \frac{\CxGLDA(0)}{\CxGLDA(1)} ],
\end{equation}
where $c_0 = +0.827 411$, $c_1 = -0.643 560$, and
\begin{align}
	\CxGLDA(1) & = \CxLDA = - \frac{3}{2} \qty( \frac{3}{4\pi} )^{1/3},
	\\
	\CxGLDA(0) & = -\frac{4}{3} \qty(\frac{2}{\pi})^{1/3}.
\end{align}
The parameters $c_0$ and $c_1$ of the \gX~enhancement factor \eqref{eq:FxGLDA} have been obtained by fitting the exchange energies of these FUEGs for $ 1 \le L \le 10$ \alert{given by Eq.~\eqref{eq:Ex}.}
$\FxgX$ automatically fulfils the constraint given by Eq.~\eqref{eq:limGLDA}.
Moreover, because $ 1\le \FxgX  \le 1.233$, it breaks only slightly the tight Lieb-Oxford bound \cite{Lieb81, Chan99, Odashima09} $\Fx < 1.174$ derived by Perdew and coworkers for two-electron systems. \cite{Perdew14, Sun16a}
This is probably due to the non-zero curvature of these FUEGs.

\alert{Albeit very simple, the functional form \eqref{eq:FxGLDA} is an excellent fit to  Eq.~\eqref{eq:CxGLDA}.
In particular, $\FxgX$ is linear in $\alpha$ for small $\alpha$, which is in agreement with Eq.~\eqref{eq:CxGLDA}. \cite{Glomium11}
Also, Eq.~\eqref{eq:CxGLDA} should have an infinite derivative at $\alpha=1$ and approached as $\sqrt{1-\alpha} \ln(1-\alpha)$.}
Equation \eqref{eq:FxGLDA} does not behave that way. 
However, it has a marginal impact on the numerical results.

\begin{figure*}
	\includegraphics[width=0.49\linewidth]{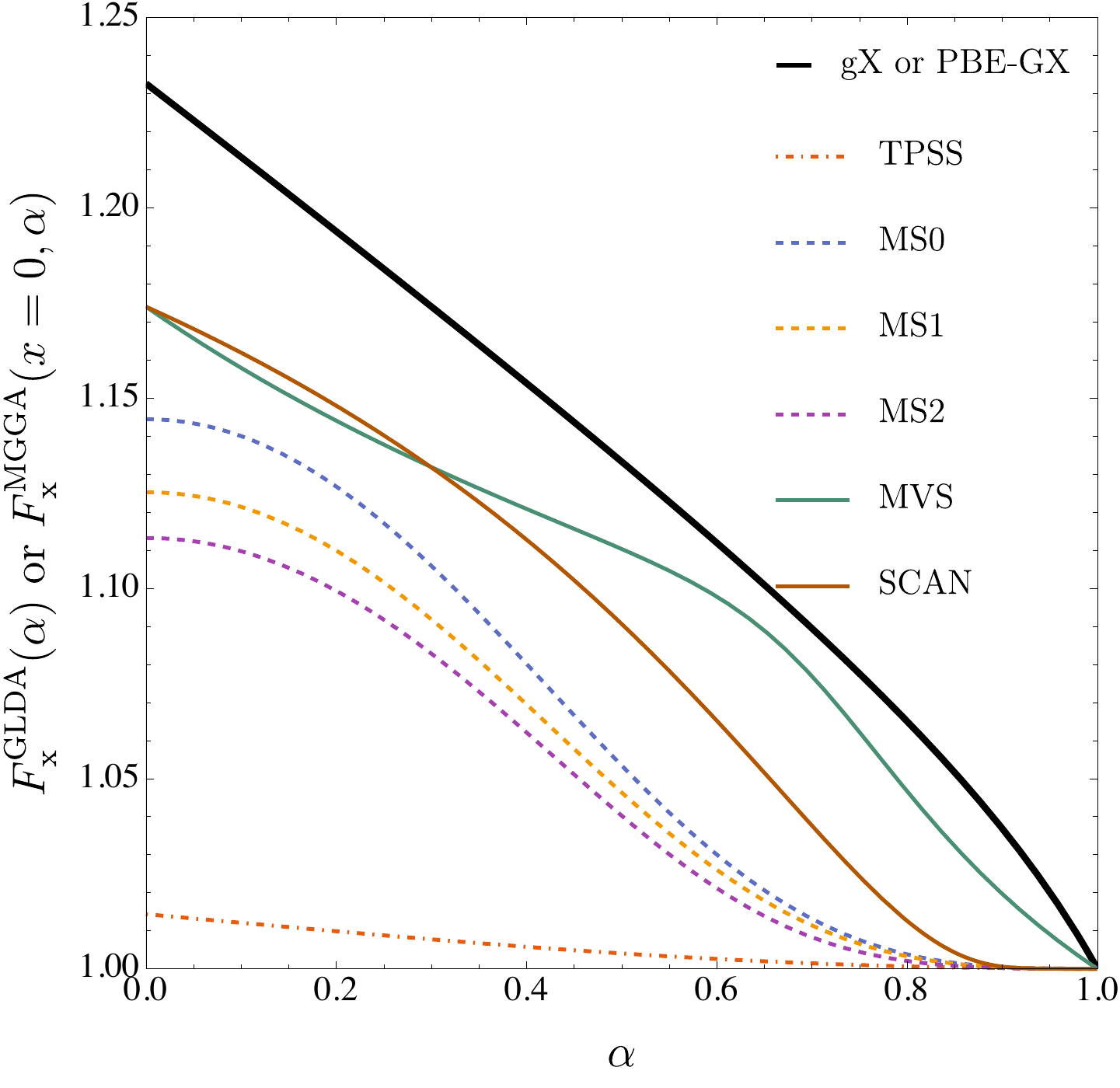}
	\includegraphics[width=0.49\linewidth]{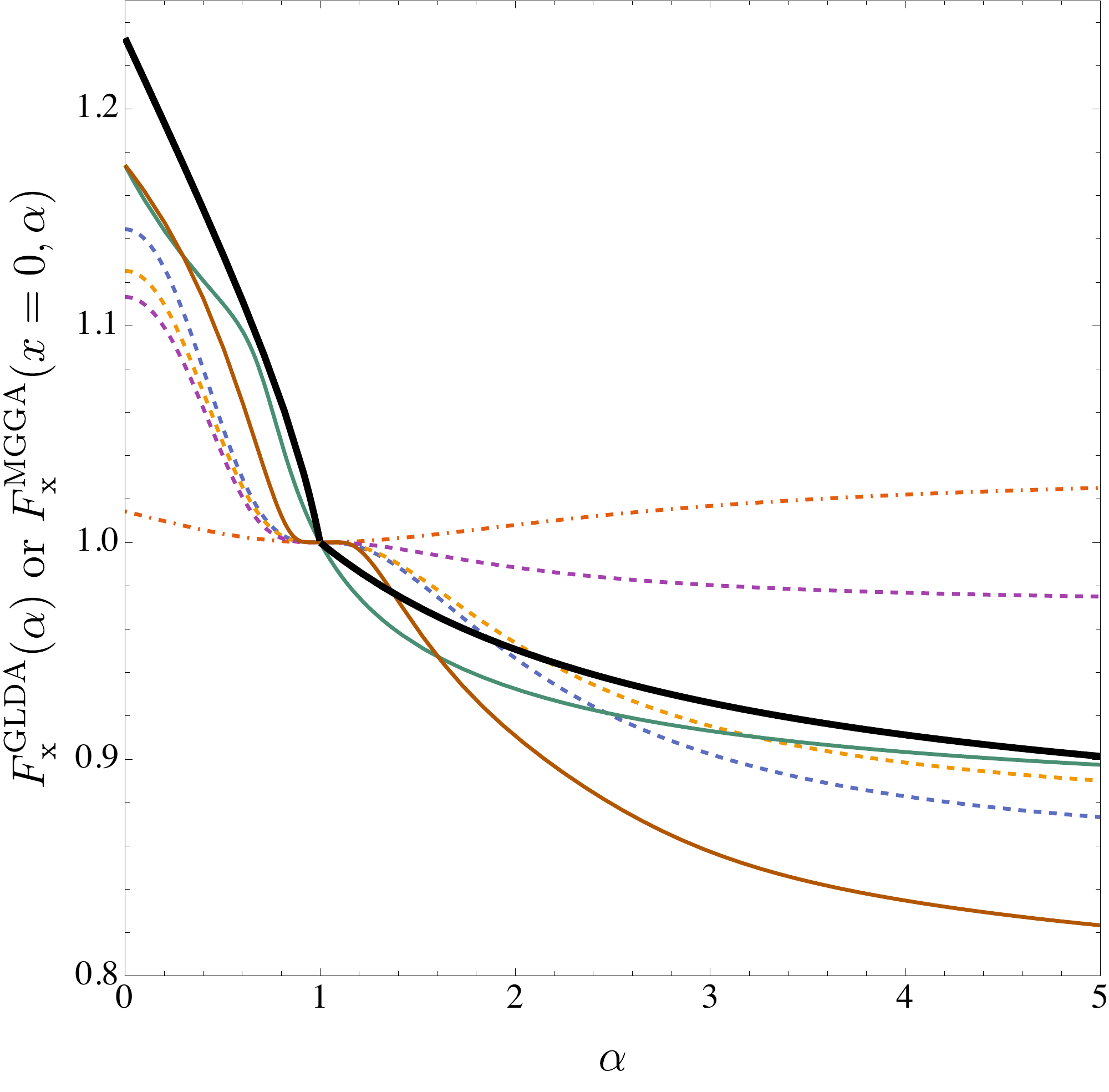}
	\caption{
	\label{fig:FxGLDA}
	\alert{Enhancement factors $\FxGLDA(\alpha)$ or $\FxMGGA(x=0,\alpha)$ as a function of $\alpha$ for various GLDA and MGGA exchange functionals.
	The TPSS functional is represented as a dot-dashed line, the MS family of functionals (MS0, MS1 and MS2) are represented as dashed lines, while the MVS and SCAN functionals are depicted with solid lines. 
	The new functionals \gX~and \PBEGX~are represented with thick black lines.
	Note that $\FxgX(\alpha) = \FxPBEGX(0,\alpha)$ for $ 0 \le \alpha \le 1$.
	For $\FxPBEGX$, $\alpha_\infty = +0.852$.}
	}
\end{figure*}

As one can see in Fig.~\ref{fig:FxGLDA}, albeit being created with FUEGs, the \gX~functional has a fairly similar form to the common MGGA functionals, such as MS0, \cite{MS0} MS1, \cite{MS1_MS2}MS2, \cite{MS1_MS2} MVS, \cite{MVS} and SCAN \cite{SCAN} for $ 0 \le \alpha \le 1$.
This is good news for DFT as it shows that we recover functionals with similar physics independently of the paradigm used to design them.
However, around $\alpha \approx 1$, the behavior of $\FxgX$ is very different from other MGGAs (except for MVS) due to the constraint of the second-order gradient expansion (which is not satisfied in our case). \cite{Ma68}
For $ 0 \le \alpha \le 1$, it is also instructive to note that the \gX~functional is an upper bound of all the MGGA functionals. 
Taking into account the inhomogeneity of the system via the introduction of $x$ should have the effect of decreasing the MGGA enhancement factor (at least for $0 \le \alpha \le 1$). 

Unlike other functionals, we follow a rather different approach and guide our functional between $\alpha=0$ and $1$ using FUEGs.
For example, the MS0 functional uses the exact exchange energies of non-interacting hydrogenic anions to construct the functional from $\alpha = 0$ to $1$, \cite{Staroverov04, MS0} while revTPSS has no constraint to guide itself for this range of $\alpha$. \cite{revTPSS}
Nonetheless, because these uniform systems only give valuable information in the range $0 \le \alpha \le 1$, we must find a different way to guide our functional for $\alpha > 1$.\alert{\footnote{\alert{Except for one- and two-electron systems, any atomic and molecular systems has region of space with $\alpha_\sigma > 1$, as discussed in details by Sun et al.\cite{Sun13a}}}}

To do so, we have extended the \gX~functional beyond $\alpha = 1$ \alert{using a simple one-parameter extrapolation:}
\begin{equation}
\label{eq:FxGMVS}
	\FxGX(\alpha) = 
		\begin{cases}
			\FxgX(\alpha),						&	0 \le \alpha \le 1,
			\\
			1 + (1-\alpha_\infty) \frac{1-\alpha}{1+\alpha},	&	\alpha > 1,
		\end{cases}
\end{equation}
where $\alpha_\infty$ is an adjustable parameter \alert{governing the value of $\FxGX$ when $\alpha \to \infty$.}
For large $\alpha$, $\FxGX$ converges to $\alpha_\infty$ as $\alpha^{-1}$, similarly to the MVS functional. \cite{MVS}
\alert{Far from claiming that this choice is optimal, we have found that the simple functional form \eqref{eq:FxGMVS} for $\alpha > 1$ yields satisfactory results (see below).}

\alert{Following the seminal work of Sham \cite{Sham71} and Kleinman \cite{Kleinman84, Antoniewicz85, Kleinman88} (see also Ref.~\onlinecite{Svendsen96}), it is also possible, using linear response theory, to derive a second-order gradient-corrected functional.
However, it does not provide any information for $\alpha >1$.}

The performance of the \GX~functional is illustrated \alert{in Table \ref{tab:atoms}}.
\alert{Although \GX~is an improvement compared to LDA, even for one- and two-electron systems, we observe that the \GX~functional cannot compete with GGAs and MGGAs in terms of accuracy.}

\begin{table*}
\caption{
\label{tab:atoms}
\alert{Reduced (i.e.~per electron) mean error (ME) and mean absolute error (MAE) (in kcal/mol) of the error (compared to UHF) in the exchange energy of the hydrogen-like ions, helium-like ions and first 18 neutral atoms for various LDA, GGA, GLDA, FMGGA and MGGA functionals.
The data for each set can be found in supplementary material.
For the hydrogen-like ions, the exact density has been used for all calculations.}}
	\begin{ruledtabular}
	\begin{tabular}{llcccccc}
			&			&	\mc{2}{c}{hydrogen-like ions}		& 	\mc{2}{c}{helium-like ions}	& 	\mc{2}{c}{neutral atoms}	\\
							\cline{3-4}						\cline{5-6}						\cline{7-8}			
			&			&	ME		&	MAE			&	ME		&	MAE			&	ME	&	MAE		\\		
	\hline
	LDA		&	D30		&	$153.5$	&	$69.7$		&	$150.6$	&	$69.5$		&	$70.3$	&	$9.1$			\\
	GGA		&	B88		&	$9.5$	&	$4.3$		&	$9.3$	&	$4.7$		&	$2.8$	&	$0.5$			\\
			&	G96		&	$4.4$	&	$2.0$		&	$4.4$	&	$2.2$		&	$2.1$	&	$0.5$			\\
			&	PW91	&	$19.4$	&	$8.8$		&	$19.1$	&	$9.3$		&	$4.5$	&	$0.8$			\\
			&	PBE		&	$22.6$	&	$10.3$		&	$22.3$	&	$10.7$		&	$7.4$	&	$0.6$			\\
	GLDA	&	\GX		&	$61.8$	&	$123.5$		&	$61.0$	&	$122.0$		&	---		&	---				\\
	FMGGA	&	MVS		&	$0.0$	&	$0.0$		&	$0.3$	&	$0.2$		&	$2.7$	&	$0.9$			\\
			&	\PBEGX	&	$0.0$	&	$0.0$		&	$0.7$	&	$0.4$		&	$1.0$	&	$1.1$			\\
	MGGA	&	M06-L	&	$44.4$	&	$88.8$		&	$12.0$	&	$24.0$		&	$4.2$	&	$2.9$			\\
			&	TPSS	&	$0.0$	&	$0.0$		&	$0.7$	&	$0.4$		&	$0.7$	&	$1.1$			\\
			&	revTPSS	&	$0.0$	&	$0.0$		&	$0.5$	&	$0.3$		&	$3.5$	&	$2.5$			\\
			&	MS0		&	$0.0$	&	$0.0$		&	$0.4$	&	$0.2$		&	$1.3$	&	$2.4$			\\
			&	SCAN	&	$0.0$	&	$0.0$		&	$0.3$	&	$0.2$		&	$1.2$	&	$1.6$			\\
	\end{tabular}
	\end{ruledtabular}
\end{table*}
\subsection{
\label{sec:FMGGA}
FMGGA exchange functionals}
One of the problem of GLDA functionals is that they cannot discriminate between homogeneous and inhomogeneous one-electron systems, for which we have $\alpha = 0$ independently of the value of the reduced gradient $x$.
\alert{For example, the \GX~functional is exact for one-electron FUEGs, while it is inaccurate for the hydrogen-like ions.}
Unfortunately, it is mathematically impossible to design a GLDA functional exact for these two types of one-electron systems.

To cure this problem, we couple the \GX~functional designed in Sec.~\ref{sec:GLDA} with a GGA enhancement factor to create a FMGGA functional (see Sec.~\ref{sec:theory}).
We have chosen a PBE-like GGA factor, i.e.
\begin{equation}
	\FxPBEGX(x,\alpha) =\Fx^\text{PBE}(x) \FxGX(\alpha),
\end{equation}
where 
\begin{equation}
\label{eq:FxPBEGX}
	\Fx^\text{PBE}(x) = \frac{1}{1+\mu\,x^2}.
\end{equation}
Similarly to various MGGAs (such as TPSS, \cite{TPSS} MVS, \cite{MVS} or SCAN \cite{SCAN}), we use the hydrogen atom as a ``norm'', and determine that \alert{$\mu = +0.001 015 549$} reproduces the exact exchange energy of the ground state of the hydrogen atom (see Sec.~\ref{sec:GLDA}). 
Also, we have found that $\alpha_\infty = +0.852$ yields excellent exchange energies for the first 18 neutral atoms.
Unlike \GX, \PBEGX~is accurate for both the (inhomogeneous) hydrogen-like ions and the (homogeneous) one-electron FUEGs, and fulfils the negativity constraint and uniform density scaling. \cite{SCAN, Perdew16}
\alert{The right graph of Fig.~\ref{fig:FxGLDA} shows the behavior of the MGGA enhancement factor for $x = 0$ as a function of $\alpha$.
Looking at the curves for $\alpha > 1$, we observe that TPSS has a peculiar enhancement factor which slowly raises as $\alpha$ increases. 
All the other functionals (including \PBEGX) decay more or less rapidly with $\alpha$. We note that \PBEGX~and MVS behave similarly $\alpha > 1$, though their functional form is different.}


\begin{figure}
	\includegraphics[width=\linewidth]{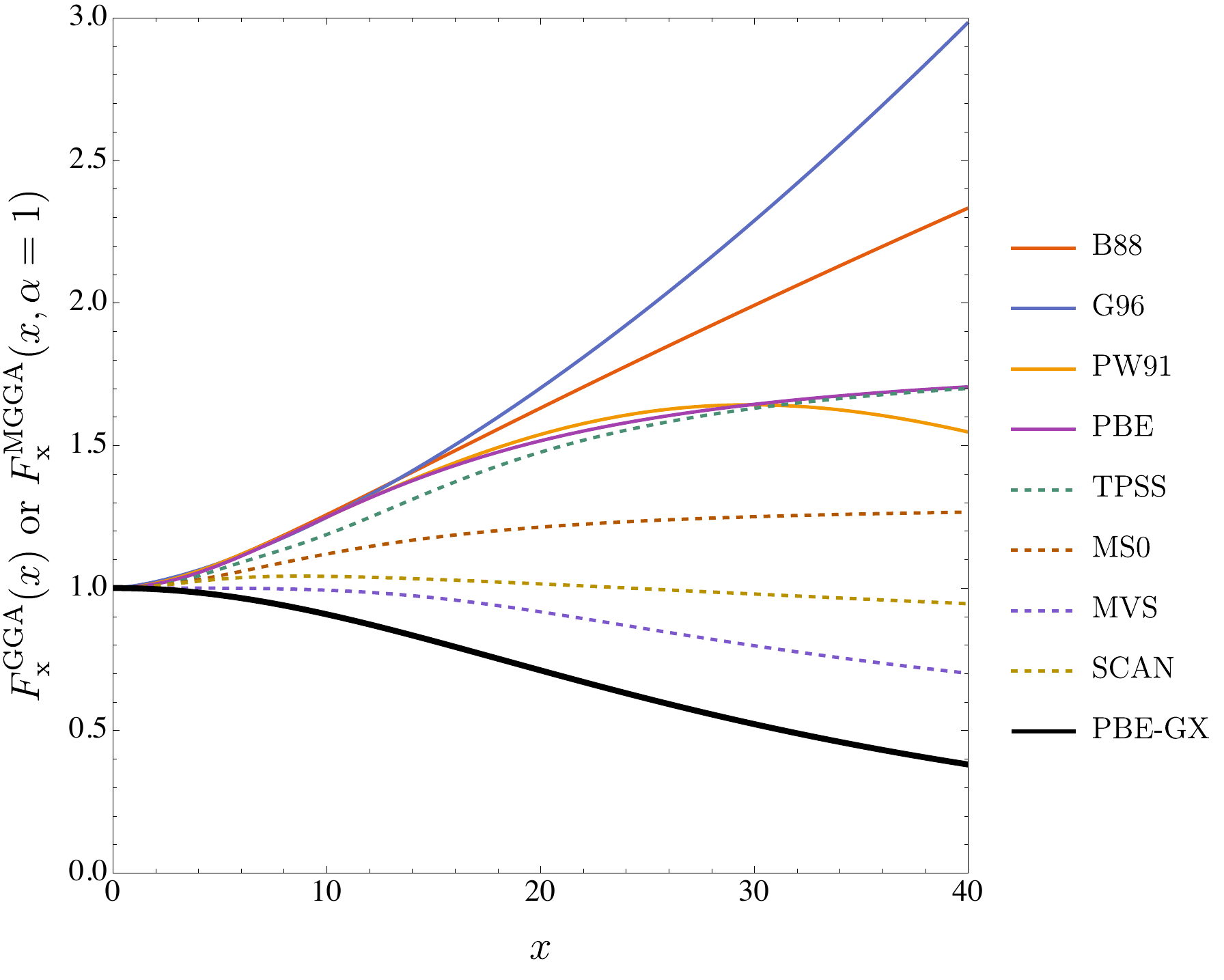}
	\caption{
	\label{fig:FxGGA}
	\alert{Enhancement factors $\FxGGA(x)$ or $\FxMGGA(x,\alpha=1)$ as a function of $x$ for various GGA, FMGGA and MGGA exchange functionals.
	The GGA functionals are represented in solid lines, while MGGAs are depicted in dashed lines.
	The new functional \PBEGX~is represented with a thick black line.}}
\end{figure}

Figure \ref{fig:FxGGA} evidences a fundamental difference between GGAs and MGGAs: while the enhancement factor of conventional GGAs does increase monotonically with $x$ and favor inhomogeneous electron densities, $\FxMGGA$ decays monotonically with respect to $x$.
This is a well-known fact: the $x$- and $\alpha$-dependence are strongly coupled, as suggested by the relationship \eqref{eq:eta-def}.
Therefore, the $x$-dependence can be sacrificed if the $\alpha$-dependence is enhanced. \cite{MS0, MVS, SCAN}
Similarly to $\FxPBEGX$, $\Fx^\text{MVS}$ and $\Fx^\text{SCAN}$ decay monotonically with $x$ (although not as fast as \PBEGX), while earlier MGGAs such as TPSS and MS0 have a slowly-increasing enhancement factor.
We have observed that one needs to use a bounded enhancement factor at large $x$ (as in Eq.~\eqref{eq:FxPBEGX}) in order to be able to converge self-consistent field (SCF) calculations.
Indeed, using an unbounded enhancement factor (as in B88 \cite{B88} or G96 \cite{G96}) yields divergent SCF KS calculations. 
\alert{Finally, we note that, unlike TPSS, \PBEGX~does not suffer from the order of limits problem. \cite{regTPSS}}

\begin{table}
\caption{
\label{tab:molecules}
\alert{Reduced (i.e.~per electron) mean error (ME) and mean absolute error (MAE) (in kcal/mol) of the error (compared to the experimental value) in the atomization energy ($E_\text{atoms} - E_\text{molecule}$) of diatomic molecules at experimental geometry for various LDA, GGA and MGGA exchange-correlation functionals.
Experimental geometries are taken from Ref.~\onlinecite{HerzbergBook}.
The data for each set can be found in supplementary material.}}
	\begin{ruledtabular}
	\begin{tabular}{lllcc}
			&	\mc{2}{c}{functional}		&	\mc{2}{c}{diatomics}			\\		
			\cline{2-3}					\cline{4-5}
			&	exchange			&	correlation		&	ME		&	MAE		\\		
	\hline
	LDA		&	D30		&	VWN5	&	$1.8$	&	$3.7$			\\
	GGA		&	B88		&	LYP		&	$0.6$	&	$1.2$			\\
			&	PBE		&	PBE		&	$0.7$	&	$1.2$			\\
	MGGA	&	M06-L	&	M06-L	&	$0.4$	&	$0.7$			\\
			&	TPSS	&	TPSS	&	$0.6$	&	$1.1$			\\
			&	revTPSS	&	revTPSS	&	$0.6$	&	$1.2$			\\
			&	MVS		&	regTPSS	&	$0.5$	&	$0.9$			\\
			&	SCAN	&	SCAN	&	$0.4$	&	$0.7$			\\
			&	\PBEGX	&	PBE		&	$0.6$	&	$1.2$			\\
			&	\PBEGX	&	regTPSS	&	$0.6$	&	$1.1$			\\
			&	\PBEGX	&	LYP		&	$0.6$	&	$1.1$			\\
			&	\PBEGX	&	TPSS	&	$0.7$	&	$1.3$			\\
			&	\PBEGX	&	revTPSS	&	$0.8$	&	$1.5$			\\
			&	\PBEGX	&	SCAN	&	$0.6$	&	$1.0$			\\
	\end{tabular}
	\end{ruledtabular}
\end{table}

\begin{figure}
	\includegraphics[width=\linewidth]{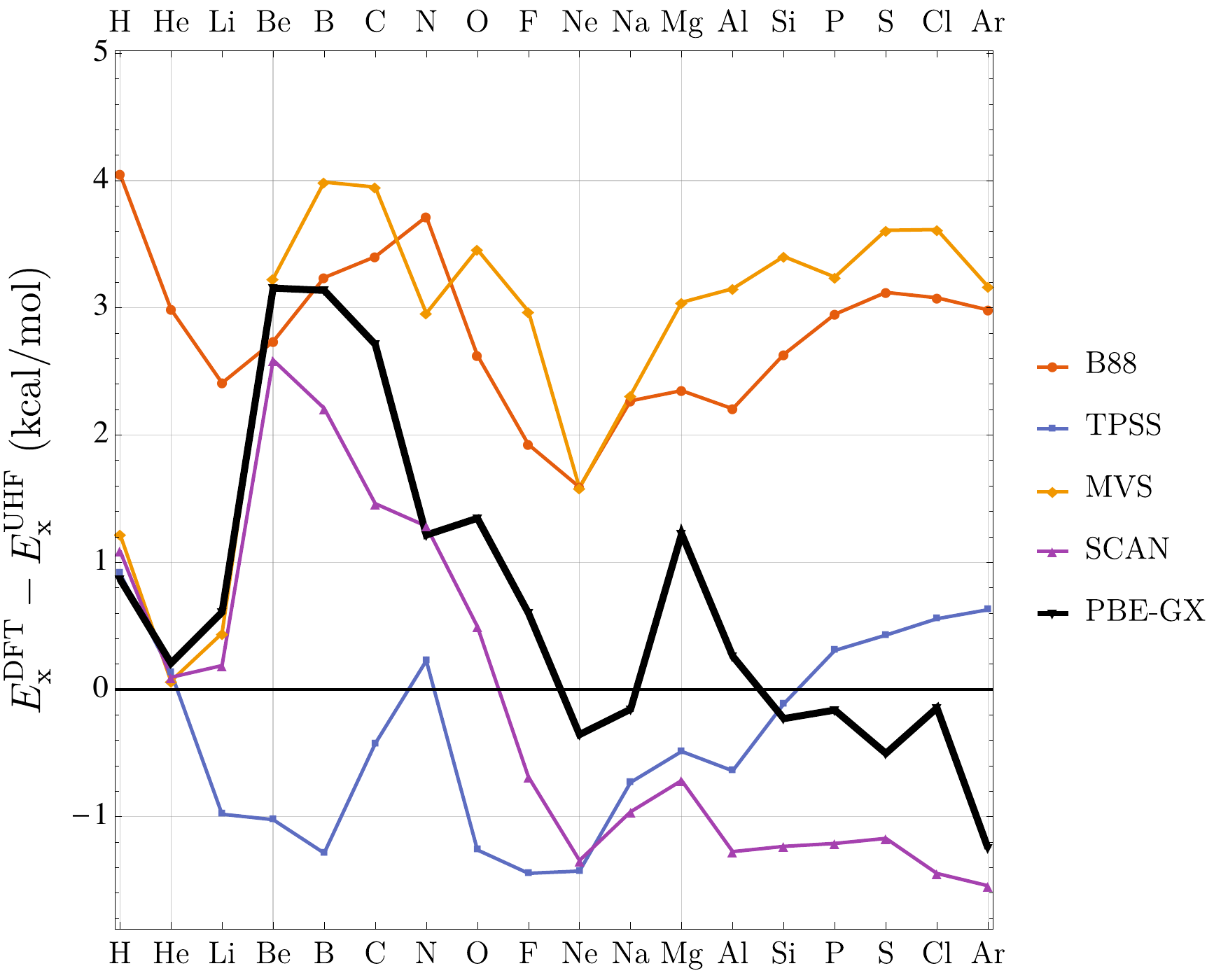}
	\caption{
	\label{fig:error}
	\alert{Reduced (i.e.~per electron) error (in kcal/mol) in atomic exchange energies of the first 18 neutral atoms of the periodic table for the B88 (red), TPSS (blue), MVS (orange), SCAN (purple) and \PBEGX~(thick black) functionals.}}
\end{figure}

\subsection{
\label{sec:resFMGGA}
How good are FMGGAs?}
\alert{The question we would like to discuss here is whether or not our new simple FMGGA functional called \PBEGX~is competitive within MGGAs.
Unlike GGAs and some of the MGGAs (like M06-L), by construction, \PBEGX~reproduces exactly the exchange energy of the hydrogen atom and the hydrogenic ions (\ce{He^+}, \ce{Li^2+}, \ldots) due to its dimensional consistency (see Table \ref{tab:atoms}).
\PBEGX~also reduces the error for the helium-like ions (\ce{H^-}, \ce{He}, \ce{Li^+}, \ldots) by one order of magnitude compared to GGAs, and matches the accuracy of MGGAs.}
For the first 18 neutral atoms (Table \ref{tab:atoms} and Fig.~\ref{fig:error}), \PBEGX~is as accurate as conventional MGGAs with a mean error (ME) and mean absolute error (MAE) of \alert{$1.0$ and $1.1$ kcal/mol}.
From the more conventional MGGAs, the TPSS and SCAN functionals are the best performers for neutral atoms with MEs of \alert{$0.7$ and $1.2$ kcal/mol, and MAEs of $1.1$ and $1.6$ kcal/mol}.
\PBEGX~lies just in-between these two MGGAs.

\alert{We now turn our attention to diatomic molecules for which errors in the atomization energy ($E_\text{atoms} - E_\text{molecule}$) are reported in Table \ref{tab:molecules} for various combinations of exchange and correlation functionals. (See the supplementary material for the list of diatomics considered in this study.)
In particular, we have coupled our new \PBEGX~exchange functional with the PBE, \cite{PBE} regTPSS \cite{regTPSS} (also called vPBEc) and LYP \cite{LYP} GGA correlation functionals, as well as the TPSS, \cite{TPSS} revTPSS \cite{revTPSS} and SCAN \cite{SCAN} MGGA correlation functionals.}

\alert{Although very lightly parametrized on atoms, \PBEGX~is also accurate for molecules.
Interestingly, the results are mostly independent of the choice of the correlation functional with MEs ranging from $0.6$ and $0.8$ kcal/mol, and MAEs from $1.0$ and $1.5$ kcal/mol.
\PBEGX~is only slightly outperformed by the SCAN functional and the highly-parametrized M06-L functional, which have both a ME of $0.4$ kcal/mol and a MAE of $0.7$ kcal/mol.}

\begin{figure*}
	\includegraphics[height=0.23\paperheight]{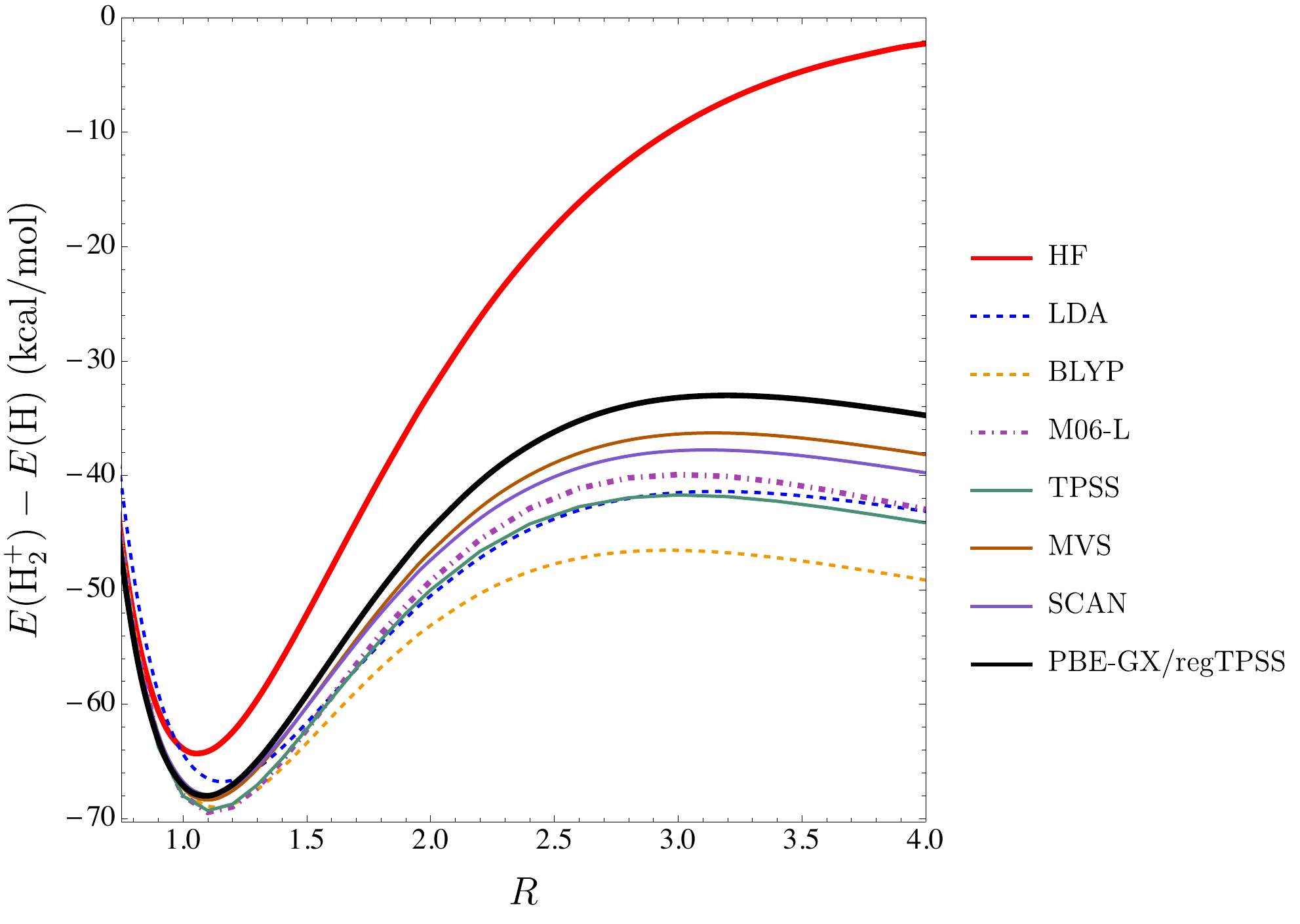}
	\includegraphics[height=0.23\paperheight]{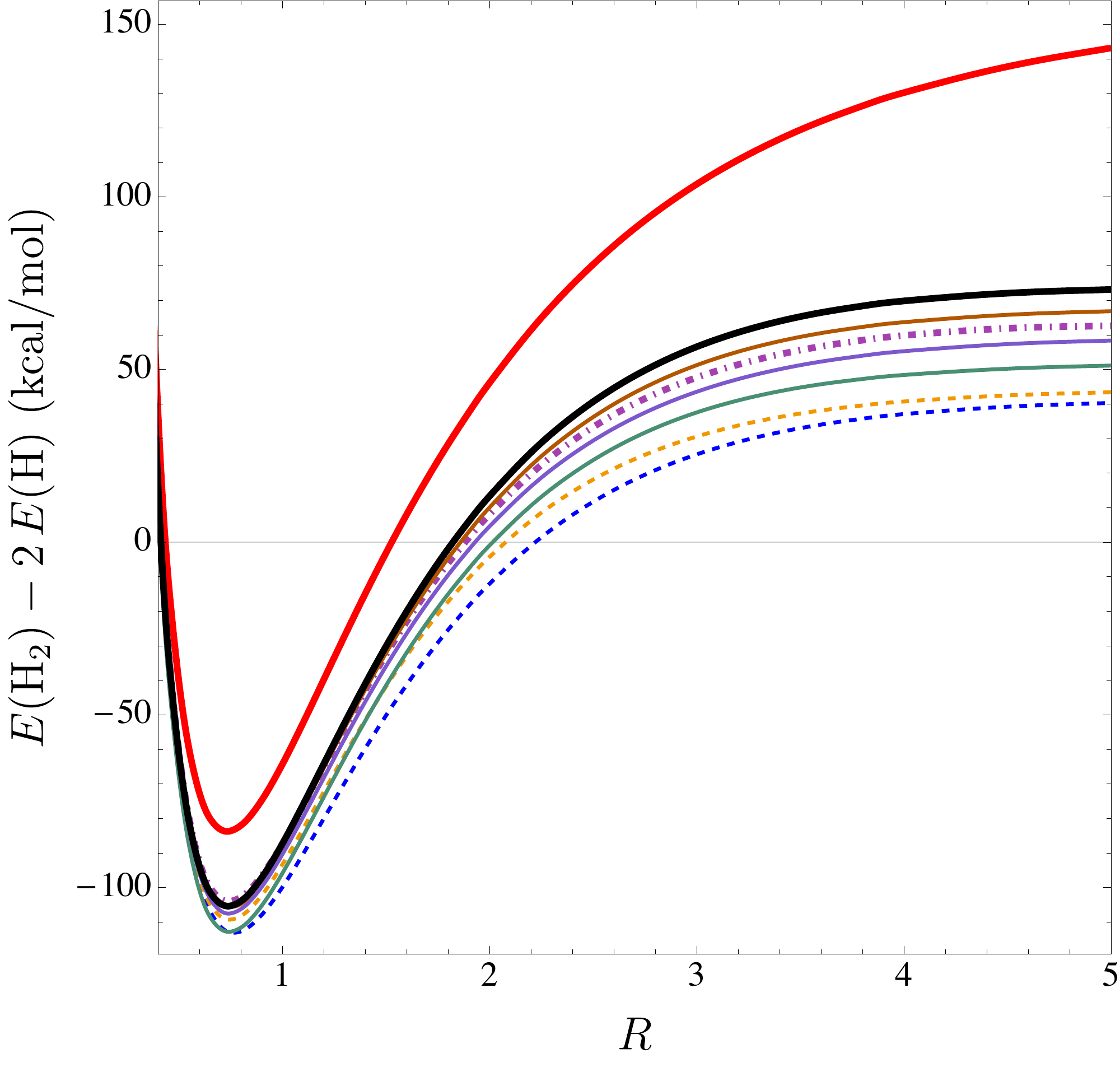}
	\caption{
	\label{fig:scan}
	\alert{Bond dissociation profile (in kcal/mol) of the \ce{H2+} (left) and \ce{H2} (right) molecules for various level of theory.
	Calculations on \ce{H2} have been performed in the restricted formalism.}}
\end{figure*}

\alert{As commonly reported, density functional approximations suffer from the self-interaction error (SIE), \cite{Merkle92, Savin96, Perdew97, Zhang98, Cohen12} i.e.~the unphysical interaction of an electron with itself.
This phenomenon is also known as the delocalization error and can be understood as the tendency of approximate functionals to artificially spread the electron density. \cite{Cohen08a, Cohen08b, MoriSanchez08, Cohen12}
To study SIE, we have reported the bond dissociation profile of the \ce{H2+} molecule for various exchange-correlation functionals (see left graph of Fig.~\ref{fig:scan}).
Obviously, HF is SIE-free. 
Thus, we can measure the SIE in a given functional as the difference between the HF dissociation curve and the curve obtained with this functional.
We see that \PBEGX/regTPSS has the smallest SIE for large bond length. However, it is still pretty significant but SIE correction could be applied (see, for example, Ref.~\onlinecite{Tsuneda14} and references therein).
Note that very similar curves are obtained by combining \PBEGX~with other correlation functionals.}

\alert{Another common pitfall of approximate density functionals known as the static correlation error \cite{Cohen08a, Cohen08c, Cohen12} can be revealed by the apparently simple problem of stretching \ce{H2}.
In the right graph of Fig.~\ref{fig:scan}, we have reported the dissociation energy profile of the \ce{H2} molecule  calculated in the restricted formalism for the same functionals.
At large bond length, the dissociation energy should reach zero but most of the functionals (including HF) do not. 
We observe that, although having the smallest SIE,  \PBEGX/regTPSS has the largest delocalization error amongst density functionals. 
Its value is still much smaller than in HF, and similar to the error in conventional MGGAs, such as MVS and M06-L.}

\section{
\label{sec:res}
Conclusion}
The purpose of the present paper is not to report an exhaustive benchmarking study but to present a new paradigm to design exchange-correlation functionals within DFT.
Using \textit{finite} UEGs (FUEGs), we have created a \textit{generalized} LDA (GLDA) exchange functional which only depends on the curvature of the Fermi hole $\alpha$.
We have also combined our newly-designed GLDA functional with a PBE-type GGA functional to create a new type of MGGAs that we have called \textit{factorizable} MGGAs (FMGGAs).
We will thoroughly investigate the  performance of our new MGGA functional in a forthcoming paper where a proper benchmarking is going to be performed.
\alert{The functional reported in the present study cannot catch dispersion interactions. 
Although special care has to be taken, \cite{Lacks93, Zhang97, Pernal09} it can be coupled with dispersion-corrected functionals. \cite{Dion04, Tkatchenko09, Lee10, mBEEF-vdW, Sun16b, Brandenburg16, Peng16}}
Also, the same approach can be applied to correlation functionals, and we will also report results on this soon.

\section*{
\label{sec:supp}
Supplementary material}
\alert{See supplementary material for raw data of Tables \ref{tab:atoms} and \ref{tab:molecules}.}

\begin{acknowledgements}
The author would like to thank Dr Andrew Gilbert for his assistance during the implementation of the present functionals in the Q-Chem package, and the University of Canterbury for a Erskine fellowship during the construction of this manuscript.
He also thanks Narbe Mardirossian for useful discussions, the Australian Research Council for a Discovery Early Career Researcher Award (DE130101441) and a Discovery Project grant (DP140104071), and the National Computational Infrastructure for generous grants of supercomputer time.
\end{acknowledgements}

\end{document}